\documentclass[prb,superscriptaddress,reprint]{revtex4-1}
\usepackage{amsmath}
\usepackage{fullpage}
\usepackage{graphicx}
\usepackage{subfiles} 
\usepackage[bottom]{footmisc}
\usepackage{comment}
\usepackage{array} 

\usepackage[utf8]{inputenc}
\usepackage[usenames,dvipsnames]{xcolor}
\usepackage{breakurl}
\usepackage[pdftex,plainpages=false,colorlinks=true,breaklinks]{hyperref}
\usepackage[toc,page,title,titletoc,header]{appendix} 
\definecolor{linkcolor}{rgb}{.8,0,0}
\definecolor{urlcolor}{rgb}{0,0,.7}
\definecolor{citecolor}{rgb}{0,.5,0}
\definecolor{acrocolor}{rgb}{0,0,.7}
\hypersetup{bookmarksopen,colorlinks=true}
\hypersetup{pdfstartview=FitH}
\hypersetup{linktocpage=true,bookmarksnumbered=true}
\hypersetup{plainpages=false,breaklinks=true}
\hypersetup{linkcolor=linkcolor,citecolor=citecolor,urlcolor=urlcolor}
\usepackage{upgreek}
\usepackage{subfigure}
\usepackage{soul}

\usepackage{xpatch} 
\makeatletter      
\xpatchcmd\@collaboration@present{(}{\medskip}{}{} 
\xpatchcmd\@collaboration@present{)}{}{}{}

\begin{document}
\title{An Interactive Gravitational-Wave Detector Model for Museums and Fairs}

\author{S.~J.~Cooper}
\affiliation{%
 Institute for Gravitational Wave Astronomy and School of Physics and Astronomy, University of Birmingham, Birmingham, B15 2TT, United Kingdom
}%

\author{A.~C.~Green}
\affiliation{%
 Institute for Gravitational Wave Astronomy and School of Physics and Astronomy, University of Birmingham, Birmingham, B15 2TT, United Kingdom
}%
\affiliation{%
Department of Physics, University of Florida, Gainesville, FL 32611, USA
}%

\author{H.~R.~Middleton}
\affiliation{%
 Institute for Gravitational Wave Astronomy and School of Physics and Astronomy, University of Birmingham, Birmingham, B15 2TT, United Kingdom
}%
\affiliation{School of Physics, University of Melbourne, Parkville, Vic, 3010, Australia}%
\affiliation{OzGrav, Australian Research Council Centre of Excellence for Gravitational Wave Discovery}

\author{C.~P.~L.~Berry}
\affiliation{%
 Institute for Gravitational Wave Astronomy and School of Physics and Astronomy, University of Birmingham, Birmingham, B15 2TT, United Kingdom
}%
\affiliation{%
 Center for Interdisciplinary Exploration and Research in Astrophysics (CIERA), Department of Physics and Astronomy, Northwestern University, 1800 Sherman Avenue, Evanston, IL 60201, USA
}%
\affiliation{%
 SUPA, School of Physics and Astronomy, University of Glasgow, Glasgow G12 8QQ, United Kingdom
}%

\collaboration{ }

\author{R.~Buscicchio}
\affiliation{Institute for Gravitational Wave Astronomy and School of Physics and Astronomy, University of Birmingham, Birmingham, B15 2TT, United Kingdom
}

\author{E.~Butler}
\affiliation{Institute for Gravitational Wave Astronomy and School of Physics and Astronomy, University of Birmingham, Birmingham, B15 2TT, United Kingdom
}

\author{C.~J.~Collins}
\affiliation{Institute for Gravitational Wave Astronomy and School of Physics and Astronomy, University of Birmingham, Birmingham, B15 2TT, United Kingdom
}

\author{C.~Gettings}
\affiliation{Institute for Gravitational Wave Astronomy and School of Physics and Astronomy, University of Birmingham, Birmingham, B15 2TT, United Kingdom
}

\author{D.~Hoyland}
\affiliation{%
 Institute for Gravitational Wave Astronomy and School of Physics and Astronomy, University of Birmingham, Birmingham, B15 2TT, United Kingdom
}%

\author{A.~W.~Jones}
\affiliation{Institute for Gravitational Wave Astronomy and School of Physics and Astronomy, University of Birmingham, Birmingham, B15 2TT, United Kingdom
}
\affiliation{OzGrav, University of Western Australia, Crawley, Western Australia, Australia}
\affiliation{OzGrav, Australian Research Council Centre of Excellence for Gravitational Wave Discovery}
\author{J.~H.~Lindon}
\affiliation{Institute for Gravitational Wave Astronomy and School of Physics and Astronomy, University of Birmingham, Birmingham, B15 2TT, United Kingdom
}
\affiliation{
Elementary Particle Physics Group, School of Physics and Astronomy, University of Birmingham, Birmingham, B15 2TT, United Kingdom}

\author{I.~Romero-Shaw}
\affiliation{%
 Institute for Gravitational Wave Astronomy and School of Physics and Astronomy, University of Birmingham, Birmingham, B15 2TT, United Kingdom
}%
\affiliation{Monash Centre for Astrophysics, School of Physics and Astronomy, Monash University, VIC 3800, Australia}
\affiliation{OzGrav, Australian Research Council Centre of Excellence for Gravitational Wave Discovery}

\author{S.~P.~Stevenson}
\affiliation{%
 Institute for Gravitational Wave Astronomy and School of Physics and Astronomy, University of Birmingham, Birmingham, B15 2TT, United Kingdom
}%
\affiliation{Centre for Astrophysics and Supercomputing, Swinburne University of Technology, Hawthorn, 3122, Victoria, Australia}
\affiliation{OzGrav, Australian Research Council Centre of Excellence for Gravitational Wave Discovery}

\author{E.~P.~Takeva}
\affiliation{%
 Institute for Gravitational Wave Astronomy and School of Physics and Astronomy, University of Birmingham, Birmingham, B15 2TT, United Kingdom
}
\affiliation{
Department of Physics and Astronomy, University of Edinburgh, James Clerk Maxwell Building, Peter Guthrie Tait Road, Edinburgh, EH9 3FD, United Kingdom}

\author{S.~Vinciguerra}
\affiliation{Institute for Gravitational Wave Astronomy and School of Physics and Astronomy, University of Birmingham, Birmingham, B15 2TT, United Kingdom
}
\affiliation{Max Planck Institute for Gravitational Physics (Albert Einstein Institute) Callinstraße 38, 30167 Hannover, Germany
}
\affiliation{Leibniz Universität Hannover Welfengarten 1, 30167 Hannover, Germany}

\author{A.~Vecchio}
\affiliation{%
 Institute for Gravitational Wave Astronomy and School of Physics and Astronomy, University of Birmingham, Birmingham, B15 2TT, United Kingdom
}
\author{C.~M.~Mow-Lowry}
\affiliation{%
 Institute for Gravitational Wave Astronomy and School of Physics and Astronomy, University of Birmingham, Birmingham, B15 2TT, United Kingdom
}%

\author{A.~Freise}
\affiliation{%
 Institute for Gravitational Wave Astronomy and School of Physics and Astronomy, University of Birmingham, Birmingham, B15 2TT, United Kingdom
}%
\affiliation{$^3$Department of Physics and Astronomy, VU Amsterdam, De Boelelaan 1081, 1081, HV, Amsterdam, The Netherlands}
\affiliation{$^4$Nikhef, Science Park 105, 1098, XG Amsterdam, The Netherlands}

\begin{abstract}

    In 2015 the first observation of gravitational waves marked a breakthrough in astrophysics, and in technological research and development. 
    The discovery of a gravitational-wave signal from the collision of two black holes, a billion light-years away, received considerable interest from the media and public. 
    We describe the development of a purpose-built exhibit explaining this new area of research to a general audience. 
    The core element of the exhibit is a working Michelson interferometer: a scaled-down version of the key technology used in gravitational-wave detectors. 
    The Michelson interferometer is integrated into a hands-on exhibit, which allows for user interaction and simulated gravitational-wave observations. 
    An interactive display provides a self-guided explanation of gravitational-wave related topics through video, animation, images and text. 
    We detail the hardware and software used to create the exhibit, and discuss two installation variants: an independent learning experience in a museum setting (the Thinktank Birmingham Science Museum), and a science-festival with the presence of expert guides (the 2017 Royal Society Summer Science Exhibition). 
    We assess audience reception in these two settings, describe the improvements we have made given this information, and discuss future public-engagement projects resulting from this work. 
    The exhibit is found to be effective in communicating the new and unfamiliar field of gravitational-wave research to general audiences. 
    An accompanying website provides parts lists and information for others to build their own version of this exhibit.
    
\end{abstract}
\maketitle

\section{Introduction and Overview}
\label{sec:introAndOverview}

Gravitational waves are ripples in space and time first predicted as a consequence of the general theory of relativity by Einstein in 1916~\cite{Einstein:1916cc}. 
A century later, and after decades of technological development, 
gravitational waves were first observed on 
14 September 2015~\cite{GW150914}. 
The signal came from two black holes orbiting each other a billion light-years from Earth~\cite{TheLIGOScientific:2016wfe,GWTC1:2019}.
The black holes merged together, creating a new bigger black hole.
The gravitational waves produced by this event spread out across the Universe, eventually reaching the Earth, where their miniscule effect was detected by the Laser Interferometer Gravitational-Wave Observatory (LIGO)~\cite{AdvancedLIGO15}. 
LIGO has since been joined in observing gravitational waves by Virgo~\cite{TheVirgo:2014hva}, and future detectors are planned with  
KAGRA~\cite{KAGRA:2013}, LIGO-India~\cite{LIGOIndia:2011} and more. %
The current global gravitational-wave detector network has made many new observations~\cite{TheLIGOScientific:2016qqj,GW151226,GW170104,Abbott:2017gyy,GW170814,GW170817,GWTC1:2019,Abbott:2020uma,LIGOScientific:2020stg,Abbott:2020tfl,Abbott:2020khf}: the beginning of a new kind of astronomy.

In anticipation of increased media coverage and public interest in gravitational-wave astronomy brought about by the first detections, the Education and Public Outreach (EPO) group of the LIGO Scientific Collaboration has worked to develop resources and activities aimed at informing and inspiring the general public, prospective students, and the wider scientific community about our work~\cite{ligoOutreach, virgoOutreach}. 
Our University of Birmingham group has a strong history of involvement with public engagement with research~\cite{laserlabs,carbone12b,gwgroup}.
Here, we describe our work developing an interactive model of a gravitational-wave detector designed to demonstrate the key technologies that have enabled gravitational-wave astronomy, and to introduce the public to this new field of astronomy. 

Museum and science-fair exhibits are an effective way of increasing interest in science~\cite{doi:10.1002/sce.21116,doi:10.1080/21548455.2019.1624991,doi:10.1080/21548455.2011.629455} and raising awareness of scientific concepts~\cite{doi:10.1002/sce.21535,doi:10.1080/21548455.2013.818260,FalkDierking2000}. 
Visits to science museums have been shown to improve long-term science knowledge~\cite{doi:10.1002/tea.20394} and adult memories of school field trips can often recall something learnt during their childhood experience~\cite{doi:10.1111/j.2151-6952.1997.tb01304.x}. 
Gravitational-wave exhibits have been developed in the past~\cite{CavagliaExhibit:2009}, and we again demonstrate their impact in engaging the public in this area of science. 

We have created an interactive exhibit that can be used both when an expert is present to explain it and as a stand-alone, non-facilitated piece which a member of the public can use to learn independently. 
Our exhibit teaches the public about gravitational waves, how they have been detected, and the kinds of astrophysical events which can be observed using them. 
The resulting piece is a long-term installation at the Thinktank Birmingham Science Museum (the Thinktank) and was featured at the 2017 Royal Society Summer Science Exhibition (RSSE). 

In this article we provide a detailed description of the design and implementation of our exhibit. 
The distinguishing features of our exhibit are summarised in Section~\ref{sec:coreDesign}. Section~\ref{sec:technicalDesign} is a more technical overview of the hardware and software used. 
Sections~\ref{sec:museumInstallation} and~\ref{sec:rsInstall} describe use of the exhibit in a museum and at a science fair respectively. 
In Section~\ref{sec:impact} we discuss the impact of our exhibit, measured through surveys as well as anecdotal examples of the public reception, and finally look to the future of these activities in Section~\ref{sec:conclusionsAndFuture}. 

Further details are provided in the appendices and accompanying website, \href{http://www.sr.bham.ac.uk/exhibit/}{www.sr.bham.ac.uk/exhibit}~\cite{michWebsite,gwexhibitsoftware}, for those who are interested in building their own model gravitational-wave detector.

\section{Design Considerations}
\label{sec:coreDesign}
Gravitational-wave science involves wide extremes of scale in the Universe. 
The colliding black holes and neutron stars that we observe have masses many times the mass of the Sun and can be billions of light-years from Earth.
However, the resulting gravitational waves arriving at Earth create minuscule changes in distance: a typical black hole collision distorts distances across the LIGO instruments by less than a thousandth of the width of a proton. 
To detect such changes, high-precision instrumentation is required, and on a large scale: each detector site is several square kilometres. 
It can therefore be challenging to communicate the science of gravitational waves in a human-relatable way.

Our exhibit is a model gravitational-wave observatory, demonstrating the core technology of current detectors like LIGO: the \emph{Michelson interferometer}~\cite{Freise:2009sf}. 
This optical configuration is often used to measure changes in distance (Section~\ref{sec:technicalDesign}).
The exhibit highlights both the behaviour of gravitational waves---changing relative distances on a small scale---and the technologies necessary to measure this behaviour. 
Such an interferometer is a common item in the tool-kit of gravitational-wave education and outreach and is often used in undergraduate laboratory experiments~\cite{ThorLabsIFO,NikhefIFO,UgoliniEtAlLIGOInTheUndergraduateLab:2019,FoxEtAl:1999}.

The exhibit design is driven by three main aims: (a) present gravitational-wave topics and concepts so that they are accessible for a broad audience, (b) attract interest in the exhibit using an appealing and exciting design~\cite{doi:10.1111/j.2151-6952.1984.tb01278.x,doi:10.1002/sce.3730790503}, and (c) be suitable for use in both a museum and a science-fair setting. 
In a museum, an exhibit needs to work as a stand-alone piece, whilst at fairs it is accompanied by experts to guide a visitor through the demonstration and answer any questions.
Exhibiting at the Thinktank allows us to engage our local community, highlighting the activities taking place in the city of Birmingham (Section~\ref{sec:museumInstallation}).
The RSSE was a national science fair providing us an opportunity to work in collaboration with several other universities (Section~\ref{sec:rsInstall}). 

The audiences in both settings are typically non-scientists with an interest in science. 
We engage our audience by pitching the exhibit material to the right level, enabling them to build upon their current understanding~\cite{doi:10.1002/tea.10070,doi:10.1080/21548455.2013.818260,doi:10.1002/sce.20014}, and conveying the subject in an interactive~\cite{doi:10.1111/j.2151-6952.2004.tb00116.x,doi:10.1002/tea.21419,doi:10.1111/j.2151-6952.1986.tb01442.x}, varied and fun way~\cite{doi:10.1111/cura.12267,doi:10.1002/tea.10068}.

Our initial design considerations were the size and weight of the model, how the unsupervised public would interact with the exhibit, and how demonstrators could use the exhibit at fairs~\cite{FalkDierking2013}. 
We wanted the gravitational-wave detector model to be as large as feasibly possible: aesthetically we wanted something eye-catching for the public~\cite{doi:10.1002/sce.3730790503}, and practically the larger size allows the components to be more easily viewed. 
At the same time, the model needed to be small enough to be easily transported for events and to and from the museum.

We settled on a $60~\mathrm{cm}$-diameter aluminium base, which can be easily lifted by two people and fits into a compact car.  
The circular shape also allows for people to easily gather around the model~\cite{doi:10.1111/cura.12267,FalkDierking2013}. 
Previous gravitational-wave exhibits have used a photodiode and speaker as an output source \cite{CavagliaExhibit:2009}, though in our initial testing this could cause positive feedback, whereby the speaker sound would further drive the mirrors. The approach of using a large interference pattern and direct interaction through software was preferred due to installation constraints. 

The gravitational-wave detector model can be used on its own, or in combination with screens and buttons that visitors can use to interact with the model and learn more about it~\cite{doi:10.1080/09500690701494050,doi:10.1002/tea.10068}. 
We have developed custom exhibit software which can be adapted to suit a specific audience and the particular interactive configuration in use.

At the Thinktank museum, our exhibit is located in a gallery containing several unrelated science exhibits, and needed to work with the museum's existing infrastructure.  
Our software and hardware needed to be durable to cope with high usage and to operate independently without maintenance for extended periods of time. 
Users guide themselves through each exhibit with the help of multimedia material;
the information presented needed to be self-explanatory, suitable for a range of interaction times, and use a range of information delivery options (video, images, text).

At the RSSE our model formed part of a collection of gravitational-wave related demonstrations, which were continuously staffed by a team of $10$ demonstrators. 
The nature of a science fair requires a short setup time--assembly needed to be simple and efficient--and the model needed to produce responses that aid the demonstrators' explanations.

In both settings we use high aesthetic and technological novelty to attract visitor attention~\cite{doi:10.1002/sce.3730790503,doi:10.1080/21548455.2013.875238,doi:10.1002/tea.10068}.

\section{Technical Design}
\label{sec:technicalDesign}
\subsection{Hardware}
\label{sec:directInteraction}

The key component of gravitational-wave detectors like LIGO and Virgo 
is the Michelson interferometer~\cite{Freise:2009sf}.
Our model gravitational-wave detector is a working Michelson interferometer.
It compares the paths of two laser light beams to detect changes in distance. 
While it cannot detect gravitational waves, it can pick up vibrations in the room, even when there is no apparent disturbances to the exhibit. 
This provides an intuitive means to illustrate that these instruments are capable of sensing distortions imperceptible to humans. 

\begin{figure}[h!]
\begin{center}
\includegraphics[width=0.5\textwidth]{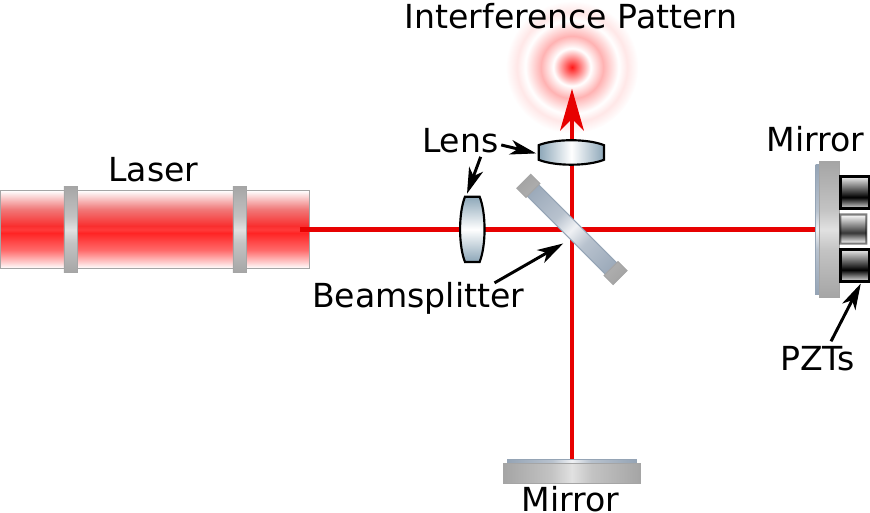}
\caption{Schematic optical layout of a Michelson Interferometer. A beamsplitter is used to split laser light equally into two perpendicular directions. Each beam reflects off a mirror, and the two beams recombine again at the beamsplitter. The interference pattern appearing at the output of the interferometer depends on differences between the two paths taken by the two beams. Piezoelectric transducers (PZTs) are used to precisely move one of the mirrors, changing the interference pattern to simulate an observation of the gravitational wave.}
\label{fig:Mich}
\end{center}
\end{figure}

The core components of a Michelson interferometer are a laser, two mirrors and a beamsplitter as illustrated in Fig.~\ref{fig:Mich}. 
The laser beam first hits the beamsplitter, where it is split in two. 
The beams travel in two arms at $90^{\circ}$ to each other to mirrors at the ends of each arm. 
After reflecting from the mirrors, the light from the two arms recombines at the beamsplitter where the two beams interfere. 

The resulting light hits a screen where it can be viewed. 
The recombined light produces a changing \emph{interference pattern}: changing distances in the detector cause the output of the detector to vary from dark to bright. 

\begin{figure}
\begin{center}
\includegraphics[width=0.5\textwidth]{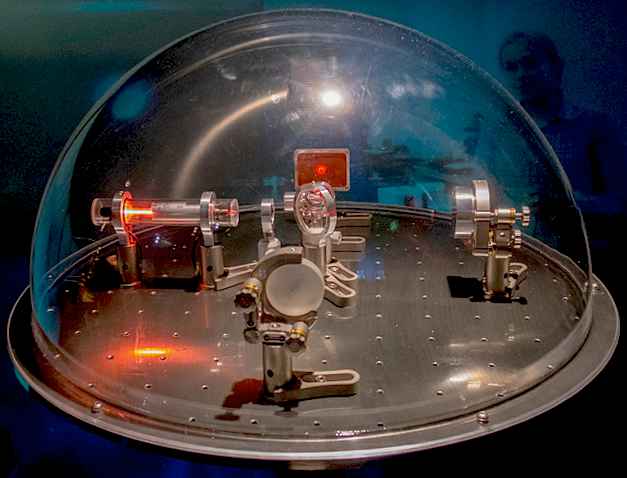}
\caption{The Michelson interferometer at the core of our exhibit, shown in situ at the Thinktank Birmingham Science Museum.}
\label{fig:MichPhoto}
\end{center}
\end{figure}

This basic interferometer setup can be built with all grades of components, from inexpensive craft mirrors and laser pointers suitable for classes of students~\cite{LIGOIFOGlue,LIGOIFOMagnets} to lab-grade optics. 
The majority of the optical components we use (beamsplitter, mirrors, mounts, etc.)\ are either lab-grade parts, or bespoke parts manufactured by the University of Birmingham's mechanical workshop. The accompanying website~\cite{michWebsite} lists example parts and variants at other price-points. 
Our cost at the time of construction was \pounds\,3600. 
High-quality components ensure the long-term stability of the configuration, achieves an appealing shiny aesthetic, and allows the public to encounter equipment which is frequently used in research laboratories.
Large mirrors and a large beamsplitter (all $2$-inch diameter) are used to increase their visibility and emphasise their importance.
Our final interferometer is shown in Fig.~\ref{fig:MichPhoto} with the laser on the left. 
In addition to the components shown in Fig.~\ref{fig:Mich}, a screen is used to project the interference pattern, webcam and photodiode are used to record the pattern and its intensity respectively.

We use a class 1 helium--neon (He--Ne) laser with an exposed view of the glowing gas. 
The exposed laser both attracts audiences and emphasises the light source. 
All optics are encased inside an acrylic dome, protecting the optical components from damage and misalignment. 
The Computer Aided Design (CAD) for the full bespoke laser mounting is shown in Fig.~\ref{fig:LaserCAD}. 
An overview of key health and safety considerations are provided in Appendix~\ref{app:HnS}.

\begin{figure}
\begin{center}
\includegraphics[width=0.5\textwidth]{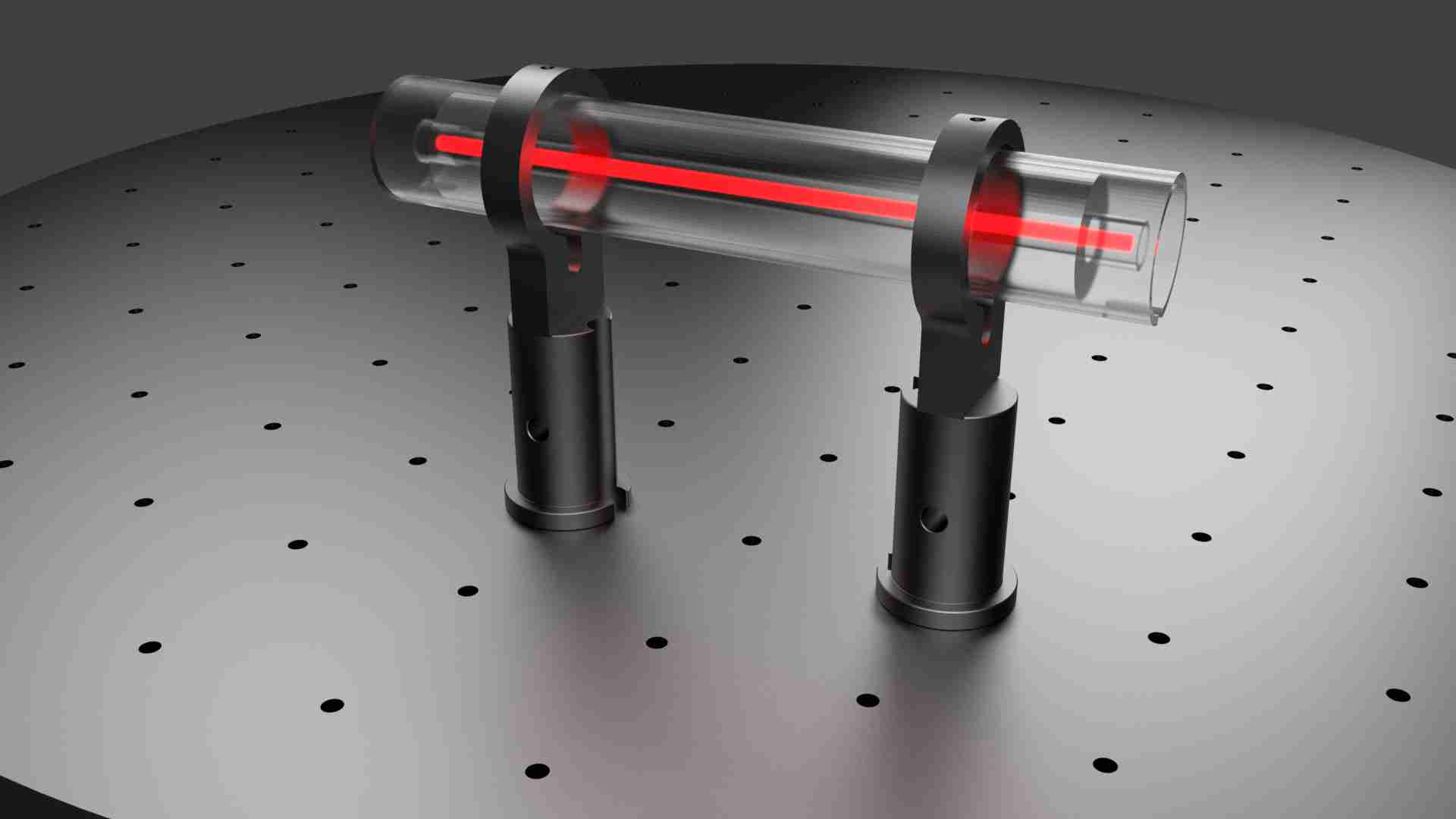}
\caption{Computer aided design used to develop bespoke parts for the exhibit. The exposed He--Ne laser tube was housed inside an insulating acrylic tube and mounted on adapted lab-grade posts. This image was rendered using Blender~\cite{blender}.}
\label{fig:LaserCAD}
\end{center}
\end{figure}

\begin{figure}
\begin{center}
\includegraphics[width=0.5\textwidth]{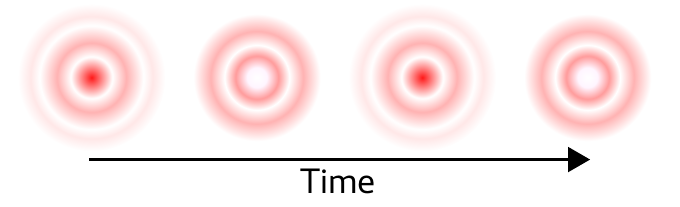}
\caption{Illustration of the ring interference pattern produced by our interferometer. When the interferometer is disturbed, the resulting concentric rings of the interference pattern change over time (left to right in the illustration) from light to dark and back again.}
\label{fig:interferencePattern}
\end{center}
\end{figure}

LIGO is fine-tuned and controlled so that the interference pattern is almost completely dark unless a gravitational wave passes through the detector. 
Our interferometer reacts to any kind of shaking motion, meaning that the interference pattern constantly flickers. 
To observe the interference pattern we tested a large variety of screen materials. The brightest, highest contrast pattern was observed using red card. 
The interference pattern in LIGO is a single spot of light changing over time. 
To demonstrate the changes more clearly to our audience, we used two diverging lenses (focal length $-50~\mathrm{mm}$) to create a large ($\sim 3~\mathrm{cm}$) beam-spot.
These, combined with a small difference in the interferometer's arm lengths, produce a ringed interference pattern that
changes depending on the relative length of the arms.
When there is no change in distance, these rings remain still, but if there is some disturbance to the interferometer the rings will expand and contract. 
By viewing the interference pattern, the audience can build an understanding that this expansion or contraction corresponds to the detector sensing a change in distance. 
Fig.~\ref{fig:interferencePattern} gives an example of how the rings' motion makes both the direction and the magnitude of the motion visually apparent, while maintaining a LIGO-like operation. 

Our Michelson interferometer can be used as a stand-alone piece. 
In this configuration, it is well suited for use with small groups under the guidance of a trained demonstrator, enabling more direct engagement~\cite{doi:10.1111/j.2151-6952.1986.tb01442.x,doi:10.1080/21548455.2019.1646439}. 
For example, the outer dome can be removed and properties of the interferometer explored, such as demonstrating alignment of optics, or pushing on the base to bend it slightly, changing the relative arm lengths.

Without a demonstrator present, it is difficult for the public to interact with the interferometer in a meaningful and safe way. 
Custom electronics and software were developed to allow the user to learn more about the exhibit, as well as interact directly with the interferometer without the need to physically touch it. 

The user can push their choice of four buttons to \emph{Send a gravitational-wave signal} 
to the interferometer, mimicking the response of a detector to a real gravitational-wave signal. 
Arcade-style buttons were chosen for their colourful and rugged nature; the bold design makes them easy to identify, attractive to children and suitable for  prolonged use~\cite{doi:10.1111/j.2151-6952.2004.tb00117.x,doi:10.1111/j.2151-6952.2011.00110.x,doi:10.1111/j.2151-6952.1997.tb01302.x,doi:10.1111/cura.12268}. 
The interaction mechanism uses three piezoelectric transducers (PZTs) mounted behind one of the end mirrors in an equilateral triangle to minimise the tilt of the mirror. 
These convert electrical voltage into motion, moving the mirror by up to $2~\upmu\mathrm{m}$.
The chosen signal is sent from a Raspberry Pi~\cite{RaspberryPi:online} to an Arduino Uno\cite{Arduino:online}, which transfers the signal to the PZTs, resulting in a moving interference pattern. 

The four predefined signals are amplified and simplified versions of the kinds of waveforms that LIGO and Virgo are searching for. 
Two are chirp signals, similar to those observed so far from merging black holes and neutron stars~\cite{GWTC1:2019}. 
The others are as-yet-unseen signals: a continuous-wave signal which is expected from rotating neutron stars~\cite[e.g.,][]{RilesCWReview:2017}, and a burst signal which could come from events like supernova explosions~\cite[e.g.,][]{SNSearch:2016}. 

To make our signals visible, they are over a trillion times larger in amplitude that the real signals. 
The signals observed by LIGO--Virgo can have frequencies over several hundred hertz, leading to changes in the interference pattern too rapid for the human eye to follow; to avoid this, we have scaled our signals to be below $25~\mathrm{Hz}$.
The simulated signals, while not exact replicas of the real events, produce visually different interference patterns, allowing members of the public to see the connection between an astrophysical object and the interference pattern observed in a LIGO-like detector. 

The ability to interact with an exhibit stimulates members of the public to ask more in-depth questions on the nature of the exhibit and surrounding themes~\cite{doi:10.1080/21548455.2019.1646439,doi:10.1002/tea.21419}. 
At science fairs, jumping or walking near our exhibit provoked questions about how to remove seismic noise from the detector output.
This created an opportunity for demonstrators to talk about different noise sources in the detector. 
As the output signal is not perfect, it also encouraged questions about the data analysis involved to identify and characterise astrophysical sources.

\subsection{Software}
\label{sec:techSoftware}\label{sec:techSoft}

Museum exhibits need to be self-explanatory~\cite{FalkDierking2013}: typically a specialist will not be present to guide the user or answer any questions. 
We developed our own exhibit software, based on a number of open-source libraries~\cite{revealJS,socketio,jonny5}, 

 to present a mixture of live data from the interferometer, video, images, and text, providing a varied range of learning materials~\cite{doi:10.1002/sce.3730700504}. 
 It has a flexible feature-set that can be used with up to two display screens, is touch-screen compatible, and can be updated with additional content as new research results emerge to provide new content for the repeat visitor~\cite{FalkDierking2013}. 
A top-level menu can be used to provide a selection of topics in sub-pages with further information, animations, a quiz, and minute-long videos. A custom graphing library was developed to show the photodiode signal directly in the exhibit software alongside the webcam. 
This allows users to direct their own learning~\cite{doi:10.1002/tea.10068,doi:10.1111/j.2151-6952.1997.tb01302.x}, and potentially build upon the understanding achieved at a previous visit~\cite{FalkDierking2000}. %
 More details of the software implementation are provided in Appendix~\ref{append:software}.

\section{Long-term Installation at the Thinktank Birmingham Science Museum}
\label{sec:museumInstallation}

The Thinktank houses a wide variety of objects and exhibits, ranging from natural sciences including fossils and wildlife specimens to science and industry including a planetarium and a large collection of steam engines
~\cite{TTBSMurl}. 
It receives $230,000$ visitors per year, including $45,000$ from schools, $10,000$ from other school-aged groups, and $152,000$ general visitors, $95$\% of whom visit with children~\cite{LaurenEmail}.

\begin{figure*}
\centering

\subfigure[\ Photograph]{\label{fig:TTp}\includegraphics[height=2.5in]{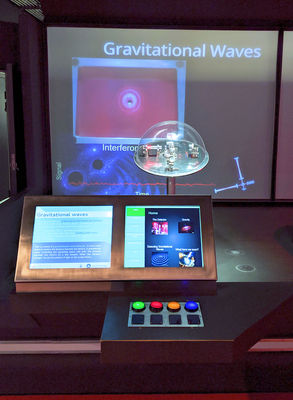}}
~
\subfigure[\ Sketch]{\label{fig:TTs}\includegraphics[height=2.5in]{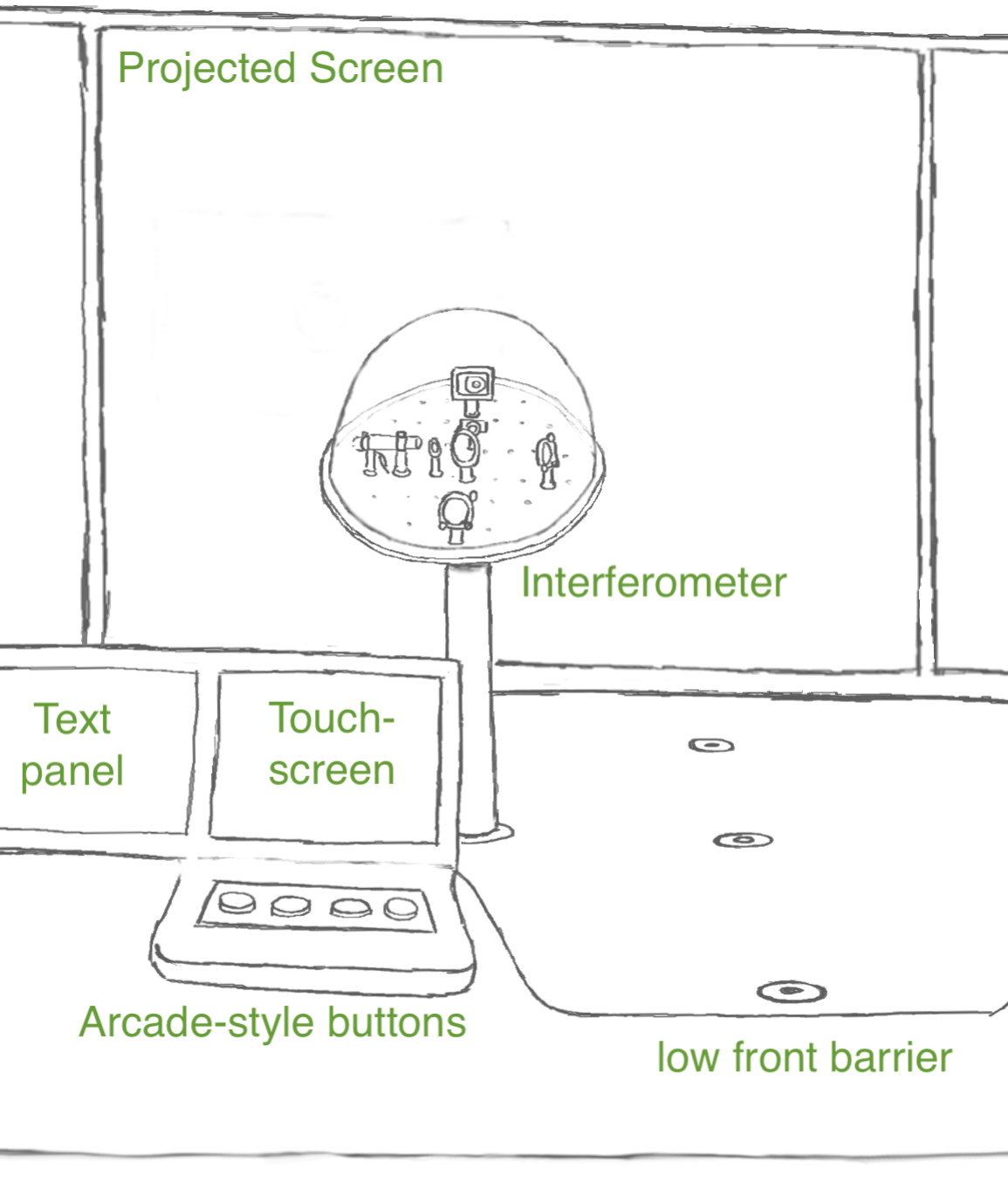}}
\caption{Our exhibit as configured for use in the Thinktank Futures Gallery.
Visitors stand next to a touchscreen panel which they can use to select from a range of topics and either read, watch videos, or take a quiz. 
Pushing the arcade-style buttons sends a signal to the driven mirror, emulating the effect of different types of gravitational-wave signals.
The interferometer model itself is mounted at an angle on a narrow post behind a low barrier approximately $1~\mathrm{m}$ from the viewer. 
All cables to and from the model, touch-screen, and buttons are concealed inside the post and under the raised false floor.
Approximately $2~\mathrm{m}$ from the viewer, at the back of the exhibit, a second projected screen displays a live feed of the interference pattern alongside the currently selected video. 
The static text panel to the left contains some overview information about the exhibit.}
\label{fig:TTconfig}
\end{figure*}

Our exhibit, initially installed in June 2016~\cite{AnnaGreenLIGOMag:2016,ThinktankInstallationBirminghamNews:2016}, is housed in the Futures Gallery. 
This gallery consists of a series of large (approximately $2~\mathrm{m} \times 3~\mathrm{m}$) bays, each focused around a central prop, as depicted for our exhibit in Fig.~\ref{fig:TTconfig}.
Props are mounted on a narrow post connected to a raised false floor which curves up to form a low (approximately $0.4~\mathrm{m}$ high) barrier between the prop and the public, and allows cabling to be fed to any part of the exhibit.
A large back-projected screen at the rear and smaller, interactive computer screen attached to the front barrier provide additional multimedia content.
Sound is played via an overhead directional speaker.
Finally, a static text panel provides an overview of each exhibit, and ensures that some information is available in the event of a technical fault. 

The Michelson interferometer itself was positioned to be ergonomically suited to the core audience of the Thinktank~\cite{TTGuidelines, Ex4All}. 
It is  $1.2~\mathrm{m}$ high, tilted $25^\circ$ forward, and located in front of the rear projected screen, which is predominantly used to display video content. 

One of our main considerations when translating our experience with hands-on demonstrations to a longstanding museum piece was developing a robust design. 
The hardware is sturdy and components are encased, to limit wear and tear so that it does not require regular maintenance. 
The electronics and computers were built to withstand the power cycling in the museum at the end of each day, and
the software was designed to use packages that will not depreciate or are easily upgraded.

Accessibility of information is also important:
it is impractical for a museum exhibit to require an expert to explain it. Therefore, we found alternate ways to convey our enthusiasm for the subject.

The back-projected screen constantly shows a large live video feed of the interference pattern, allowing visitors to clearly see  
how the pattern changes due to floor vibrations as they move around, or as they send a mock signal via the arcade buttons. 
A simultaneous view of the actual pattern on the screen inside the interferometer dome, along with real-time graphing of the center of the interference pattern, shows that this video is a live feed from the instrument they see in front of them.

In a museum visitors will often spend limited time interacting with any particular exhibit. 
It is important to make the scientific information quick and easy to access, whilst also providing depth and variety for longer interactions~\cite{FalkDierking2013}. 
The software described in Section~\ref{sec:techSoft} was developed specifically to achieve this, and to enable direct interaction with the exhibit. 

The content was developed in collaboration with the Thinktank to ensure the language was suitable and accessible for both children and adults. 
The colour scheme was checked to be colour-blind friendly using Color Oracle~\cite{ColorOracleurl} and the fonts chosen were sans-serif as there are indications that these fonts may be more dyslexia-friendly~\cite{BDA1url,BDA2url}; 

video material is subtitled and audio is played through directional overhead speakers.

There are two points of user interaction in the museum implementation of the exhibit: a touch-screen computer, and the four buttons used to trigger exaggerated examples of gravitational-wave-like signals. 
The buttons can be pressed at any time during interaction with the exhibit.
The touch-screen provides a selection of topics and a quiz.

Typically, selecting a topic on the touch-screen will trigger a short (under $90~\mathrm{s}$) video to play on the rear screen, during which visitors are free to click through several short ($<80$ words per page) text~\cite{doi:10.1111/j.2151-6952.1997.tb01302.x,FalkDierking2000} and graphic pages that expand on the content of the video. 
The video duration is displayed on the screen, so that a viewer is aware that the video is short~\cite{doi:10.1080/00220671.2013.833010}. 

The videos cover four core topics: 
(a) an explanation of the gravitational-wave detector model on display and what the interference pattern means;
(b) the nature of gravity and how gravitational waves are produced; 
(c) how interferometers are used to search for gravitational waves; 
(d) the detection of gravitational waves, including the first detection in 2015 and other observations since.  
Each video features enthusiastic members of our group as well as video clips, graphics and animations created by others in the LIGO Scientific Collaboration~\cite{EPOMultimediaresources}. 
This is intended to help the public make a human connection to the science we discuss, and to share in our excitement for the subject~\cite{doi:10.1080/21548455.2013.875238,FalkDierking2013}. 
The four presenters (two female and two male) were all PhD students, 
chosen as we considered them to be relatable and avoided the aged-professor stereotype of a physicist~\cite{doi:10.1002/tea.21419,doi:10.1111/cura.12267,FalkDierking2000}. 

The quiz facilitates more active interaction: the user tests their knowledge and receives confirmation of their understanding~\cite{doi:10.1080/09500690701494068}. 
It is intentionally short (four questions) and displays the final score in a chalk-board image next to Einstein.

The software, interferometer and static display together enable varying levels of self-guided interaction within a museum setting. 
A short engagement allows a little information about gravitational waves to be gained; however, much more detailed information is available if a visitor chooses a prolonged interaction.

\section{The Royal Society Summer Exhibition}\label{sec:rsInstall}
The RSSE~\cite{RSSE17url} is an annual week-long festival hosted at The Royal Society in London, celebrating cutting-edge science and technology in the UK. 
It is open to all and free; 
the audience ranges from school groups and the general public to politicians and celebrities.  
Each year, around $20$ teams from universities and industry are selected to develop a science-fair-style exhibit, staffed full-time by members of each group.

In 2017, UK members of the LIGO EPO group~\footnote{LIGO Scienitifc Collaboration universities and institutions at RSSE: University of Birmingham, Imperial College London, University of Glasgow, Cardiff University, University of the West of Scotland, University of Southampton, University of Sheffield, AEI Hannover, AEI Potsdam--Golm, Milde Marketing} were selected to jointly host a stand at the RSSE named \textit{Listening to Einstein's Universe}~\cite{L2Ursse, L2Uurl}.
The $4~\mathrm{m} \times 2~\mathrm{m}$ stand used printed fabric panels to create zoned areas that visitors could walk through, interacting with activities and props representing different areas of our research. 
The University of Birmingham 
provided two key components of the stand: (a) a series of apps and games developed by the Birmingham group~\cite{carbone12b, gwoptics, laserlabs}, and (b) our gravitational-wave detector model, on loan from the Thinktank.

\begin{figure}
    \centering
\subfigure[\ Photograph]{\label{fig:RSp}\includegraphics[height=2in]{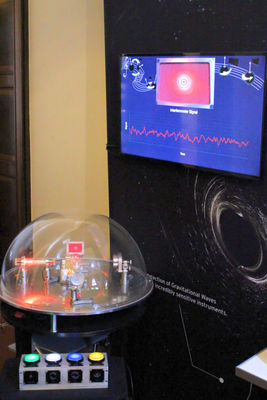}}
~
\subfigure[\ Sketch]{\label{fig:RSs}\includegraphics[height=2in]{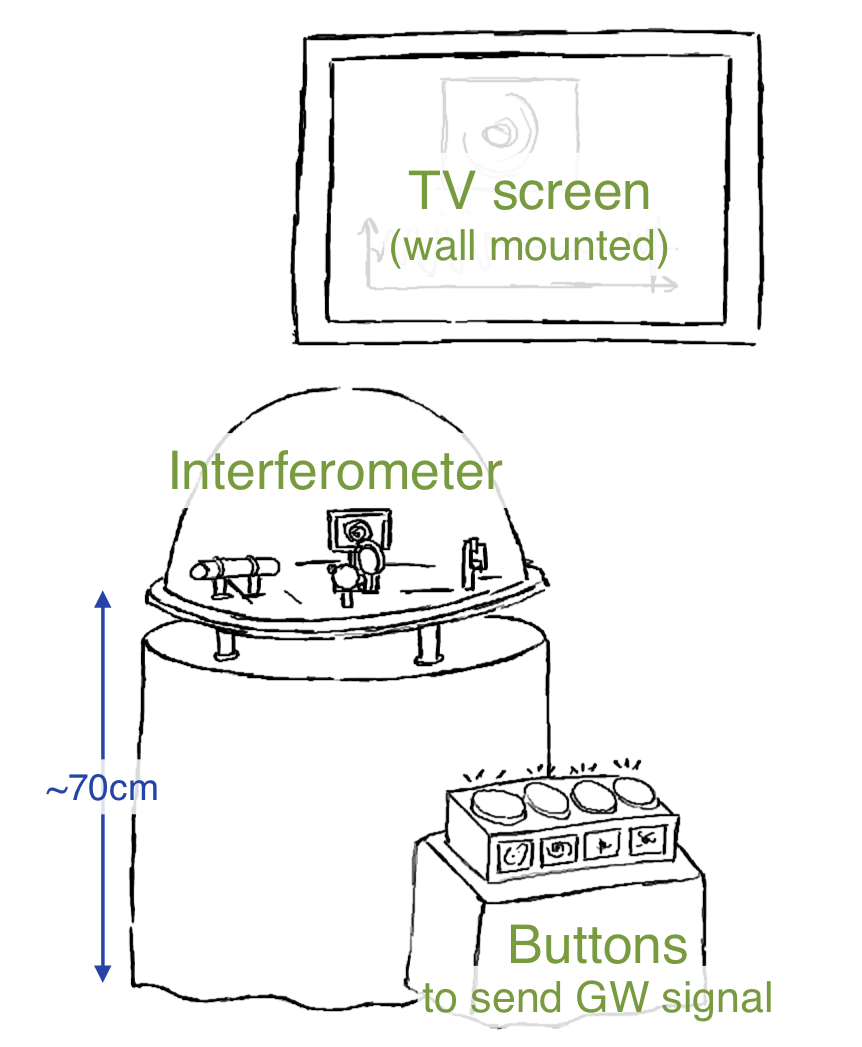}}
\caption{Our exhibit as configured for use at the RSSE 2017. 
The gravitational-wave detector model is mounted on three feet and sits on a round table $0.7~\mathrm{m}$ above the floor.
Additional electronics and resources as concealed on a lower shelf behind a table cloth. 
The arcade-style buttons are housed in a separate box on an independent low table, so that pushing a button does not shake the interferometer. 
A $40$-inch television screen is mounted into the fabric panelling of the \textit{Listening to Einstein's Universe} stand (on the upper-right), and displays live feeds of both the interference pattern, and a graph of the resulting measured signal.}
\label{fig:RSconfig}
\end{figure}
 
The hardware and software configuration used at the RSSE is depicted in Fig.~\ref{fig:RSconfig}. 
Compared to the Thinktank configuration, the core components---the interferometer and a display screen---were identical, however the self-guided content was not required as experts would always be on hand. 
Instead, a single smaller screen was used to show the live video feed of the interference pattern alongside a graph of the intensity at the center of the interference pattern. 
A simplified display allowed for tailored explanations: a short overview at times of high footfall (such as for school groups), or an expanded explanation when time and visitor interest allowed.

The robustness required for the Thinktank meant that transporting the model was relatively simple, and that the installation time and maintenance of the exhibit were minimal.
The dome on the interferometer meant that we could safely invite visitors to take an up-close look at the optics, which was not only useful for explaining the model but also a necessity in the compact space. 
The model was located towards the back of the \textit{Listening to Einstein's Universe} stand, so aesthetic choices such as the use of bright coloured lights and shiny components helped attract visitors in this setting too~\cite{FalkDierking2013}. 
The arcade buttons were housed in a separate box to prevent the physical button press shaking the exhibit and drowning out the injected signal. 

Overall, the combination of aesthetic and practical design choices made for a museum setting also resulted in a model that is effective and exciting to use at science fairs such as the 2017 RSSE.
The adaptable multimedia content in particular enables the exhibit to be tailored to a wide range of contexts.

\section{Impact}\label{sec:impact}

Since the exhibit's installation at the Thinktank in July 2016, it has spent over three years 
 on display there, as well as being included as part of the \textit{Listening to Eintein's Universe} stand at RSSE.

Each stage of the exhibit's life has enabled us to explore different aspects of user engagement and reception, as well as the broader impact of the project. 
In this section, we describe our feedback collection at the RSSE, the improvements we have implemented to the exhibit based on experiences gained in a science-fair setting, 
how observations in the museum setting led to further modifications, and consider the wider impact of the project for our group and the opportunities resulting from it.

\subsection{Feedback from the Royal Society Summer Science Exhibition 2017}
\label{sec:impactRoyalSociety}

In 2017, $10,123$ members of the public, $2002$ students and $262$ teachers visited the RSSE~\cite{RSSEdemo}.
School groups (age 14+) were required to register in advance to attend the dedicated sessions, bringing up to $25$ students per group~\cite{RSSE_2017_schoolguide}. 
This means that students at the RSSE are more likely to be from schools with proactive, well supported class teachers who were motivated to pursue out-of-classroom activities, and the students themselves are a sub-selection from their year group or class (e.g., most interested, most likely to benefit from attending, etc.), biasing the sample compared to the entire population of schoolchildren. 
The audience at the RSSE is therefore not ideal for engaging hard-to-reach demographics with research.

Feedback from $171$ exhibit visitors was collected via surveys, as detailed in Appendix~\ref{sec:question}.
The results indicate that we had a positive impact on those who attended: 
of participants who rated their interest in physics, $54\%$ replied \emph{Highly} or \emph{Very highly} interested before the visit, rising $83\%$ after.
However, the results cannot capture the long-term impact of the exhibit and our interactions with visitors; this is an area of active exploration for the group in the future.

\subsection{Feedback from the Thinktank Futures Gallery Installation}

To gauge the impact of our exhibit at the Thinktank, we monitored visitor interactions with the exhibit and conducted a survey.  
The number of survey participants was small, and thus robust conclusions could not be drawn.
However, we found useful information and several areas of improvement were clear from observations of museum visitor behaviour and interaction. 

In its first iteration, the exhibit was placed towards the back of the Futures Gallery in a relatively dark corner. 
Opposite this position was one of the museum highlights: \emph{RoboThespian}, a talking, singing robot. 
Many visitors were observed to head directly for the robot, skipping the back corner of the gallery entirely~\cite{FalkDierking2013}. 
On the day of the survey, only $13$ of $200$ people who entered the gallery interacted with our exhibit.  
In light of this, we worked with the museum to place our exhibit closer to the gallery entrance, providing both a more prominent position, where visitors are liable to spend more time~\cite{FalkDierking2013}, and also offering better lighting to attract the visitor. 

Of those who interacted with the exhibit, nine took part in the survey, and asked about their prior knowledge of gravitational waves. 
Three had not heard of gravitational waves before seeing the exhibit and none had any awareness of the University of Birmingham's involvement in the discovery of gravitational waves. 
All found the exhibit at least \emph{Quite informative} and \emph{Fairly easy to use}. 

Our observations at the Thinktank, and our experiences at the RSSE, led us to make modifications to improve upon the museum implementation, which originally could only be interacted with through a roller-ball mouse and a single click button. 
In the RSSE, we found that adding interaction through the arcade-style buttons worked well, and 
the subsequent addition of a touch screen at the Thinktank has brought the exhibit more up-to-date with modern technology and, therefore, more familiar to the visitor~\cite{doi:10.1080/09500690701494050}.

\subsection{Wider Impact}

The experience gained in designing, implementing and evaluating the exhibit, as well as exposure to science communication professionals, has enabled us to explore new means of sharing gravitational-wave science to non-expert audiences in engaging and accessible ways. 
It has also led to further work beyond the exhibit itself. 

One such project is an interdisciplinary collaboration with audio--visual digital artist Leon Trimble~\cite{LeonTrimbleurl}. 
The project, \textit{Gravity Synth}, is a musical instrument combining a Michelson interferometer with a modular synthesiser~\cite{GravitySynth}. 
The interference pattern is converted into sound via a photodiode, and processed through a modular synthesiser, exploring the relationship between gravitational waves, vibrations and sound. 
The \textit{Gravity Synth} has been performed at a variety of events with themes ranging from arts and music to science~\footnote{Events where \textit{Gravity Synth} has featured include: BBC's Digital Planet 18th birthday show (2019)~\cite{GravitySynthBBC}, Cheltenham Science Festival (2019), Lunar Festival (2018), Future Everything (2018), The Superposition (2017), Pint of Science (2017), and the Interact Engagement Symposium (2017)}.

We have also formed collaborations with other university departments who are keen to include a Michelson interferometer as part of their own public engagement schemes. 
Our group has built a similar interactive interferometer for the University of East Anglia as well as additional smaller, more portable variations for our own use. 
Our website~\cite{michWebsite} details component lists and instructions for use by others to build their own interferometer. 
Alongside this, we are investigating making low-cost interferometer kits with novelty elements such as building blocks as a more affordable and fun means for schools to create their own Michelson interferometers, similar to existing examples from the LIGO EPO group which use magnets~\cite{LIGOIFOMagnets} and glue~\cite{LIGOIFOGlue}.

\section{Conclusions}\label{sec:conclusionsAndFuture}
At this exciting time for gravitaitonal-wave astronomy, our aim for this project was to bring this research to a wider audience in an accessible way. 
We have designed and built a physical exhibit and custom-made exhibit software that are able to explain what gravitational waves are, how they are detected, and the recent discoveries.
By installing the exhibit in the Thinktank Birmingham Science Museum, we have a long-term means of increasing community awareness of the research taking place at a local university.
The exhibit has also been shown at the 2017 Royal Society Summer Exhibition, providing a geographically wider reaching impact in the shorter term. 
We have monitored the reception to our exhibit and taken action in response to what we have learnt. 

This project has led to further work in gravitational-wave public engagement, including collaborations with artists bringing this research to a potentially new audience at arts and music festivals. 
In the long term, the project will have a lasting role on an international scale with online instructions and parts lists~\cite{michWebsite} to enable others, including school groups, to build their own versions of this exhibit.

\section*{Acknowledgements}\label{sec:acknowledgements}

This project was funded by the Science and Technology Facilities Council Small Award (project number ST/N005767/1) and the Royal Astronomical Society Public Engagement Grant (2015).
The authors are grateful to both the Thinktank Birmingham Science Museum and The Royal Society for their support and advice for the exhibit development and installation. 
The exhibit videos were produced in collaboration with the communications department of the College of Engineering and Physical Sciences at the University of Birmingham. 
The authors express their deep gratitude to all those who have helped with this project since its inception, in particular: Steve Brookes and all those in the University of Birmingham mechanical workshop; John Bryant and David Stops for additional technical support; Alejandro Vigna G{\'o}mez for his contributions to the exhibit videos; Siyuan Chen and Carl-Johan Haster for useful discussions for the exhibit development, and Julia Dancu and Luke Scantlebury-Smead for conducting the Thinktank survey. 
The authors thank Laura Trouille and Joey Shapiro Key for their comments on this manuscript; the LIGO Education and Public Outreach working group for their support and for use of multimedia resources, and Sarah Cole, Kris Vogt Veggeberg and Jayatri Das for useful recommendations for literature on exhibit design. 
We are grateful to the anonymous referees for their feedback. 
This work was also supported by STFC Grant ST/N000633/1 and the University of Birmingham. 
CPLB is supported by the CIERA Board of Visitors Research Professorship, and NSF Award PHY 1912648. 
This research is supported by the Australian Research Council Centre of Excellence for Gravitational Wave Discovery (OzGrav) (project number CE170100004).
ACG is supported by NSF Awards PHY 1806461 and PHY 2012021.
This work has been assigned LIGO document number P2000036. 

\appendix
\section{Health \& Safety Considerations} 
\label{app:HnS}
Our exhibit naturally needed to follow health and safety protocols to be suitable for unsupervised use by all ages, especially curious children, when used at fairs and in the museum. 

The primary hazard is the high voltage required for the laser ($1~\mathrm{kV}$ when running). 
The laser is therefore mounted inside an acrylic tube with acrylic end-caps, and grounded to the base plate (Fig.~\ref{fig:LaserCAD}). 
All active electrical components are encased in a plastic box underneath the optical base so that these cannot be accessed.
The the acrylic dome both protects the public from the high voltage laser, and protects the optics from stray fingerprints, dust, and getting knocked out of alignment.
Eye safety is assured by using a class 1 laser, ensuring that all beams are contained in the circular base plate, and the restricted access of the dome. 

At the Thinktank, the gallery configuration means that visitors have no physical contact with the prop on display: all the interaction is made through the computer screen and buttons at the front of the exhibit.
However, the low barrier means that it is easy for the visitors, for example young children, to climb over and touch props.
As such, we made sure that the post attachment of our interferometer to the false floor was sturdy.

\section{Software Design}
\label{append:software}

There is existing software to create museum exhibits, such as Open Exhibits~\cite{openExhibits} or Intuiface~\cite{intuilab}.
A drawback of these is that they are either written in older programming languages such as Actionscript 3, as is the case with Open Exhibits, or are costly to design and run, like Intuiface. Existing solutions therefore lacked the flexibility required for our exhibit.

We have developed our own exhibit software based on a number of Javascript libraries including reveal.js, socket.io and johnny-five \cite{revealJS,socketio,jonny5}.
The advantages of using these libraries are that they: are modern; have a relatively low barrier to entry; can be visualised easily through the use of HTML and CSS, which are key web technologies; are easy to develop; and are widely supported and will be supported for years to come.
This allowed more time and effort to be spent on content creation, rather than the functionality of the exhibit.
Wide support for this software also makes installation in a variety of locations, with different sets of requirements, easier to manage. 
The packages used in this project are also open-source, reducing the overall cost.

The software is flexible in terms of the available features and customisation. 
It can be used with one or two display screens and is touch-screen compatible.
A set of navigable pages can be created to guide a user through the materials. 
The top-level menu provides a selection of topics, leading to sub-pages with further information. 
 This allows users to direct their own learning~\cite{doi:10.1002/tea.10068,doi:10.1111/j.2151-6952.1997.tb01302.x}, and potentially build upon the understanding achieved at a previous visit~\cite{FalkDierking2000}. %
 Within the sub-pages, a combination of animations, images and text can be displayed. 
 The display can also be timed to return to the home screen after a set idle time, which is desirable when the exhibit software has been left on a sub-page so that it is ready for the next user.
 Selecting a topic can also trigger a pre-recorded video to begin playing, as well as simultaneously displaying live data from the exhibit.
 
 In the gravitational-wave detector model, a webcam is used to display an enlarged view of the current interference pattern on the screen, and 
 a photodiode embedded in the centre of the screen takes a light intensity reading, which is live-plotted using custom-written graphing software. 
 This helps larger groups of people to clearly see the interference pattern from a distance and the changing intensity reading in response to different signals.

\begin{figure}
\begin{center}
	\includegraphics[width=0.49\textwidth]{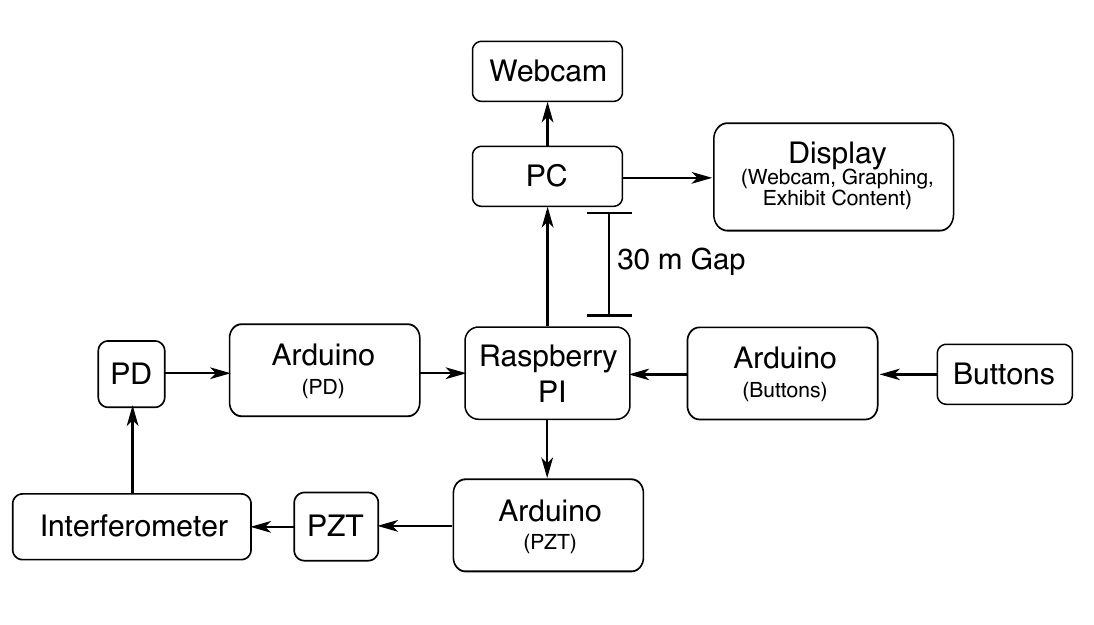}
	\caption{Signal flow chart for the exhibit. A Raspberry PI runs the exhibit software and interfaces with the piezoelectric transducers (PZTs) and the photodiode (PD). When used at the Thinktank Birmingham Science Museum, the PC is separated by $30~\mathrm{m}$ from the rest of the exhibit, so communication must be be done via ethernet rather than a simpler, faster USB connection (USB is limited to $5~\mathrm{m}$).}
	\label{fig:signalPath}
\end{center}
\end{figure}

A schematic view of all the signal paths used in our exhibit is shown in Fig.~\ref{fig:signalPath}.
As described in Section~\ref{sec:directInteraction}, the user can directly interact with the exhibit via a selection of buttons. 
Pressing a buttons simulates a gravitational-wave signal disturbing the interferometer. These are in the form of two chirping signals, representing the merger of binaries of black holes or neutron stars~\cite{Abbott:2016bqf}; a continuous wave, representing a spinning neutron star, and a burst, representing a supernova explosion. 
These signals are designed to take inspiration from their real-life counterparts, while being reshaped in duration, amplitude and frequency to be visually distinguishable four our visitors.

 The buttons can also be lit up in patterns via the software, making them attractive to push.

The specific configuration of software and hardware used in the Thinktank and the RSSE are detailed in Section~\ref{sec:museumInstallation} and Section~\ref{sec:rsInstall}, respectively.

\section{Survey Design and Results}

\label{sec:question}

We created two survey types for use at the RSSE: electronic and paper. 
The electronic survey was completed via a pair of tablets at the exhibit stand. 
This targeted individuals or small groups, typically older teenagers or adults. 
Questions primarily asked the user to rate their opinion on a Likert scale from $1$ to $5$. 

The paper survey was aimed at younger attendees by using a range of graphical question formats rather than the traditional multiple choice questionnaire. 
This format was also better suited to large groups since the paper forms could be distributed quickly to an entire class. 
Both survey types were short (eight and ten questions for the paper and electronic versions respectively),
taking no more than a few minutes to complete. 

The questions within the two surveys were not identical due to the different survey formats. However, both aimed to assess change in the individual's interest in physics and gravitational-wave research, as well as their interest in specific parts of the \textit{Listening to Eintein's Universe} stand.
The demographics of those surveyed through both formats are summarised in Table~\ref{tab:RSDemog}. 
A conscious effort was made to maintain a gender balance across each session of those surveyed over the course of the week.

\begin{table}
\begin{tabular}{r l}
\hline
\hline
\textbf{Survey respondents} & \textbf{Total (Paper $|$ Electronic)} \\
\hline
Total responses & $171$ ($63$ $|$ $102$)\\
\\
Identifying as:       & \\
               female   & $39\%$ ($34\%$ $|$ $40\%$)   \\
                 male   & $55\%$ ($52\%$ $|$ $58\%$) \\
  other/not specified   & ~$6\%$ ($14\%$ $|$ $2\%$) \\
\\
Aged 18 or under & $51\%$ ($84\%$ $|$ $31\%$) \\
\hline
\hline
\end{tabular}
\caption{\label{tab:RSDemog}
Demographic overview of survey respondents at the RSSE 2017. 
The labels Paper and Electronic signify whether the information was gathered via paper form (used mostly for younger people) or the electronic form via tablet.
}
\end{table}

\begin{table*}
\begin{tabular}{r ccc}
\hline
\hline
How much has your knowledge of gravitational waves changed? & \emph{No change} & \emph{Large increase} \\
                                                  (Electronic)  & $1\%$       & $80\%$ \\
\\
\hline
How much has your knowledge of LIGO changed? & \emph{No change} & \emph{Large increase} \\
                                    (Electronic) & $7\%$         & $66\%$ \\
\\
\hline
Rate your interest in physics before \& after visiting the RSSE$^*$ & 1-step increase & 2-step increase$^{**}$ \\
                                                                       (Paper) & $49\%$     & $13\%$     \\ 
\\

\hline
\hline
\end{tabular}
\caption{\label{tab:RSresults}
Selected survey results from the RSSE 2017. The left column shows the question where the labels Electronic and Paper are as described in Table~\ref{tab:RSDemog}. The right columns show some key results. \\
\scriptsize{
* $83\%$ of those specifying no change were already \emph{Highly} or \emph{Very highly} interested in physics before attending the RSSE. 
Most people described their interest before the RSSE as \emph{OK} ($32\%$) or \emph{High} ($32\%$), and after the RSSE as \emph{Very highly} ($51\%$).
Percentage increase in interest across genders: female $75\%$, male $48\%$, other/undisclosed $75\%$.\\
** All of those specifying a 2-step increase were aged 11--16 ($22\%$ of 14--16 year-olds and $8\%$ of 11--13 year-olds)
}}
\end{table*}

A summary of responses to key survey questions is given in Table~\ref{tab:RSresults}. 
While the surveys generally discussed the exhibit as a whole, many people spent a significant fraction of their time at the interferometer model, and, therefore, the results are considered reflective of the model gravitational-wave detector. 

According to questions from the electronic survey, we found that \textit{Listening to Eintein's Universe} was successful in both its core goal of increasing people's awareness of gravitational waves ($80\%$ \emph{Large increase}), and in informing them about LIGO ($66\%$ \emph{Large increase}), which was just one of several research projects mentioned at the exhibit. 
Similarly, the paper survey 

 indicated that $62\%$ of those asked were more interested in physics than they were previously, while a large majority of the rest were already \emph{Highly interested} in physics before they arrived. 
Further analysis was possible using the data from the paper forms. 
The exhibit was particularly effective at engaging some of the 11--16 age range: all of those indicating a \emph{Large increase} in interest in physics ($8$ of the $63$ people surveyed) were in this age bracket. 
It also effectively engaged girls, $75\%$ of whom indicated increased interest. 
It is possible that this is a result of our efforts to deliberately include both male and female volunteers in every session of the exhibition.

\bibliographystyle{myunsrt}

\end{document}